\documentclass{ws-procs9x6}

\begin{document}

\title{Pseudo + quasi SU(3): Towards a shell-model description of 
heavy deformed
nuclei\footnote{\uppercase{T}his work was supported in part by 
\uppercase{CONAC}y\uppercase{T}
(\uppercase{M}\'exico) and
the \uppercase{US N}ational \uppercase{S}cience \uppercase{F}oundation.}}

\author{Jorge~G. HIRSCH, Carlos~E. VARGAS}

\address{Instituto de Ciencias Nucleares,
Universidad Nacional Aut\'onoma de M\'exico, \\
Apartado Postal 70-543 M\'exico 04510 DF, M\'exico \\
E-mail: hirsch@nuclecu.unam.mx}

\author{Gabriela Popa}

\address{Department of Physics, Rochester Institute of Technology, \\
Rochester, NY 14623-5612, USA\
E-mail: gpopa@rochester.rr.com}

\author{Jerry~P. DRAAYER}

\address{Department of Physics and Astronomy,
Louisiana State University, \\
Baton Rouge, Louisiana 70803, U.S.A. \\
E-mail: draayer@lsu.edu}

\maketitle

\abstracts{
The pseudo-SU(3) model has been extensively used to study normal parity
bands in even-even and odd-mass heavy deformed nuclei. The use
of a realistic Hamiltonian that mixes many SU(3) irreps has allowed
for a successful description of energy spectra and electromagnetic
transition strengths. While this model is powerful, there are situations
in which the intruder states must be taken into account explicitly. The
quasi-SU(3) symmetry is  expected to complement the model, allowing for
a description of nucleons occupying normal and intruder parity orbitals
using a unified formalism. }

\section{Introduction}

The SU(3) shell model\cite{Ell58} has been successfully applied to a 
description of
the properties of light nuclei, where a harmonic oscillator mean field
and a residual quadrupole-quadrupole interaction can be used to describe
dominant features of the nuclear spectra.
However, the strong spin-orbit interaction renders the SU(3) truncation
scheme useless in heavier nuclei, while at the same time pseudo-spin
emerges as a good symmetry\cite{Hec69}.

The pseudo-SU(3) model\cite{Hec69,Rat73} has been used to describe
normal parity bands in heavy deformed nuclei. The scheme takes full
advantage of the existence of pseudo-spin symmetry, which refers to the
fact that single-particle orbitals with $j = l - 1/2$ and $j = (l-2) +
1/2$ in the $\eta$ shell lie close in energy and can therefore be
labeled as pseudo-spin doublets with quantum numbers
$\tilde j = j$, $\tilde\eta = \eta -1$ and $\tilde l = l - 1$. The origin
of this symmetry has been traced back to the relativistic Dirac
equation\cite{Blo95}.

A fully microscopic description of low-energy bands in even-even and
odd-A nuclei has been developed using the pseudo-SU(3) model\cite{Dra82}. The
first applications used pseudo-SU(3) as a dynamical symmetry, with a single
irreducible representation (irrep) of SU(3) describing the yrast band up
to the backbending region\cite{Dra82}. On the computational side, the
development of a computer code to calculate reduced matrix elements of
physical operators between different SU(3) irreps\cite{Bah94} represented a
breakthrough in the development of the pseudo-SU(3) model. With this code in
place it was possible to include symmetry breaking terms in the interaction.

Once a basic understanding of the pseudo-SU(3) model was achieved and computer
codes enabling its application developed, a powerful shell-model theory for
a description of normal parity states in heavy deformed nuclei emerged. For
example, the low-energy spectra and B(E2) and B(M1) electromagnetic transition
strengths have been described in the even-even rare earth isotopes
$^{154}$Sm, $^{156,158,160}$Gd, $^{160,162,164}$Dy and
$^{164,166,168}$Er\cite{Beu98,Beu00,Pop00,Dra01} and in the odd-mass 
$^{157}$Gd,
$^{159,161}$Tb, $^{159,163}$Dy, $^{159}$Eu, $^{161,169}$Tm, and $^{165,167}$Er
nuclei\cite{Var00a,Var00b,Var01}.

In the present contribution we review recent results
obtained using a modern version of the pseudo-SU(3) formalism, which
employs a realistic  Hamiltonian with single-particle energies plus
quadrupole-quadrupole and  monopole pairing interactions with strengths
taken from known systematics\cite{Var00a,Var00b}. Its eigenstates are
linear combinations of the coupled pseudo-SU(3) states. The quasi SU(3)
approach for intruder states is also discussed, together with
its implications regarding a unfied description of a system with nucleons
occupying normal and intruder parity orbitals.

\section{The Pseudo SU(3) Basis and Hamiltonian}

Many-particle states of $n_\alpha$ active nucleons in a given normal
parity shell $\eta_\alpha$, $\alpha = \nu$ (neutrons) or $\pi$ (protons),
can be classified by the group chain $~U(\Omega^N_\alpha )
\supset U(\Omega^N_\alpha / 2 ) \times U(2) \supset SU(3) \times SU(2)
\supset SO(3) \times SU(2) \supset SU_J(2)$, where each group in the
chain has associated with it quantum numbers that characterize its
irreps.

The most important configurations are those with highest spatial
symmetry\cite{Dra82,Var98}. This implies that $\tilde{S}_{\pi , \nu} = 0$ or
$1/2$, that is, the configurations with pseudo-spin zero for an even
number of nucleons or $1/2$ for an odd number are dominant.
In some cases, particularly for odd-mass nuclei, states with
$\tilde{S}_\pi$ = 1 and $\tilde{S}_\nu$ = ${\frac 3 2}$ must also be taken
into account, allowing for coupled proton-neutron states with total
pseudo-spin $\tilde{S}$ = ${\frac 1 2}$, ${\frac 3 2}$ or ${\frac 5 2}$.
Since pseudo-spin symmetry is close to an exact symmetry in the normal parity
sector of the space, a strong truncation of the Hilbert space can be invoked.
However,the pseudo spin-orbit partners are not exactly degenerate and this
introduces a small pseudo-spin mixing in the nuclear wave function.

The Hamiltonian,
\begin{eqnarray}
    H & = & H_{sp,\pi} + H_{sp,\nu} - \frac{1}{2}~ \chi~ \tilde Q \cdot
            \tilde Q - ~ G_\pi ~H_{pair,\pi} ~\label{eq:ham} \\
      &   & - ~G_\nu ~H_{pair,\nu} + ~a~ K_J^2~ +~ b~ J^2~ +~ A_{sym}~
            \tilde C_2, \nonumber
\end{eqnarray}

\noindent includes spherical Nilsson single-particle energies for $\pi$
and $\nu$ as well as the pairing and quadrupole-quadrupole interactions,
with their strengths taken from systematics\cite{Var01,Rin79}.
Only the parameters $a$, $b$ and $A_{sym}$ are used fit to the data.
A detailed description of each term in the Hamiltonian
(\ref{eq:ham}) can be found in Ref.\cite{Var00b}.

\begin{figure}[t]
\centerline{\epsfxsize=4.5in\epsfbox{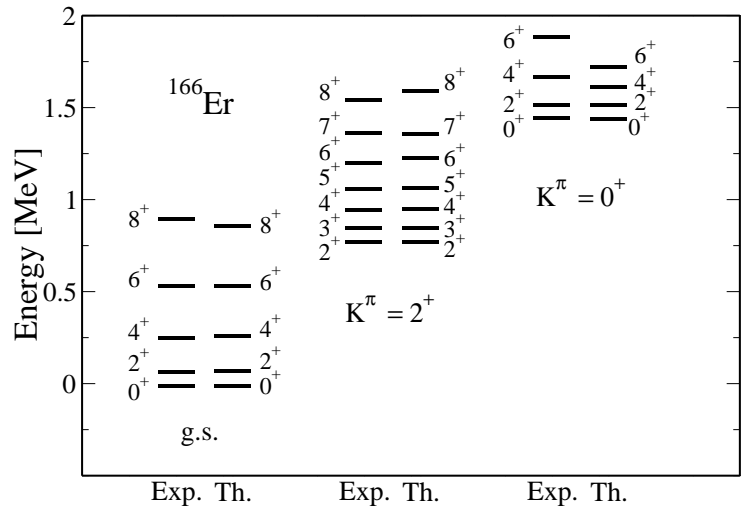}}
\vspace*{-0.9cm}
\caption{Energy spectra of $^{166}$Er \label{sp-er}}
\end{figure}

\begin{figure}[t]
\centerline{\epsfxsize=4.0in\epsfbox{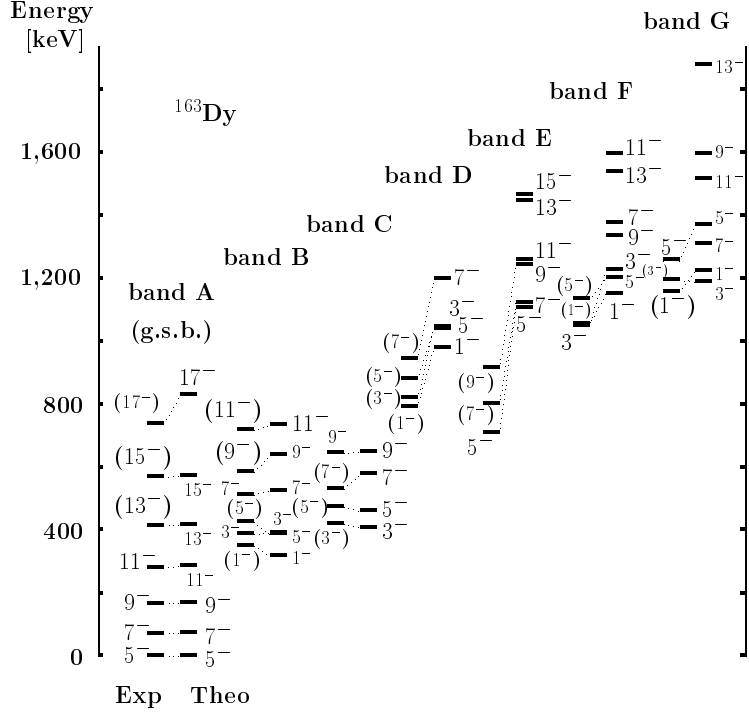}}
\vspace*{-0.4cm}
\caption{Energy spectra of $^{163}$Dy. \label{sp-dy}}
\end{figure}

The electric quadrupole operator is expressed as\cite{Dra82}

\begin{equation}
Q_\mu = e_\pi Q_\pi + e_\nu Q_\nu \approx
e_\pi {\frac {\eta_\pi +1} {\eta_\pi}} \tilde Q_\pi +
e_\nu {\frac {\eta_\nu +1} {\eta_\nu}} \tilde Q_\nu , \label{q}
\end{equation}

\noindent with effective charges $e_\pi = 2.3$ and $e_\nu =
1.3$\cite{Dra82,Beu98}. The magnetic dipole operator is

\begin{equation}
T^1_\mu = \sqrt{\frac {3}{4\pi}} \mu_N \{ g^o_\pi L^\pi_\mu +
g^S_\pi S^\pi_\mu + g^o_\nu L^\nu_\mu + g^S_\nu S^\nu_\mu \}
\label{eq:t1op}
\end{equation}

\noindent where the `quenched' $g$ factors for $\pi$ and $\nu$ are used.
To evaluate the M1 transition operator between eigenstates of
the Hamiltonian (\ref{eq:ham}), the pseudo SU(3) tensorial expansion of
the T1 operator (\ref{eq:t1op})\cite{Cas87} was employed.


\section{Some representative results}

The experimental and theoretical ground-, beta- and gamma-bands in
$^{166}$Er are shown in Fig. \ref{sp-er}. Having a close connection
with the rotor Hamiltonian, the pseudo SU(3) model is particularly well
suited to describe these bands. The term proportional to $K_J^2$ allows
the position of the gamma ($K_J = 2$) band-head to be fit, a particularly
difficult task in many fermionic models.

Experimental and theoretical B(E2) transition strengths in $^{166}$Er
  are shown in Table \ref{er166-be2}. Effective charges used are 
1.25$e$ and 2.25$e$.
Transitions between states in the ground-state band are of the order 
of $e^2b^2$,
while those from the $\gamma-$ to ground-state bands are far smaller.
The agreement with the experimental information is remarkable.

\begin{table}[h!]
\tbl{Experimental and theoretical B(E2) transition strengths for $^{166}$Er.
\vspace*{1pt}}
{\footnotesize
\begin{tabular}{|c|c|c|}
\hline
  $J_i \rightarrow J_f$ & \multicolumn{2}{|c|}{B(E2)[$e^2b^2 \times 10^{-2}$]}\\
\cline{2-3}
& Exp. & Th. \\\hline
{} &{} &{} \\[-1.5ex]
  $0_{gs} \rightarrow 2_{gs}$    &  580     $\pm$  27    & 580 \\
  $2_{gs}\rightarrow 4_{gs}$     &  303     $\pm$  20    & 299\\
  $4_{gs}\rightarrow 6_{gs}$     &  273     $\pm$  35    & 265\\
  $6_{gs}\rightarrow 8_{gs}$     &  258     $\pm$  35   & 251\\
  $2_{\gamma}\rightarrow 4_{gs}$ &  0.363  $\pm$  0.027 & 1.485\\
  $2_{\gamma}\rightarrow 2_{gs}$ &  4.915  $\pm$  0.038 & 10.310\\
  $0_{gs}\rightarrow 2_{\gamma}$ &  15     $\pm$   1    & 17 \\
  $4_{gs}\rightarrow 5_{\gamma}$ &  $\geq$ 0.27         & 3.22\\
\hline
\end{tabular}\label{er166-be2} }
\vspace*{-13pt}
\end{table}


Fig. \ref{sp-dy} shows the yrast and six excited normal parity bands in
$^{163}$Dy.  The integer numbers denote twice the angular momentum of
each state.
Experimental\cite{nndc} energies are plotted on the left-hand-side
of each column, while their theoretical values are shown in the
right-hand-side. These results should be compared with the three
bands described in an earlier study\cite{Var01}, where the same Hamiltonian and
parametrization was employed but the Hilbert space was restricted to
$\tilde{S}_\pi =$ 0 and $\tilde{S}_\nu = {\frac 1 2}$ states. The present
description reproduces almost all the data reported for normal parity
bands in this nucleus.

\begin{table}[h!]
\tbl{Theoretical B(E2) transition strengths for $^{167}$Er.
\hfill \vspace*{1pt}}
{\footnotesize
\begin{tabular}{|cc|cc|}
\hline
$J^-_i \rightarrow J^-_f$ & B(E2) &$J^-_i \rightarrow J^-_f$ & B(E2)\\ \hline
$1/2^-_1 \rightarrow 3/2^-_1$   & 275    & $5/2^-_2 \rightarrow 
7/2^-_2$ & 310 \\
$3/2^-_1 \rightarrow 5/2^-_1$   &  59    & $7/2^-_2 \rightarrow 
9/2^-_3$ & 252 \\
$5/2^-_1 \rightarrow 7/2^-_1$   &  25    & $9/2^-_3 \rightarrow 
11/2^-_4$ & 186 \\
$7/2^-_1 \rightarrow 9/2^-_1$   &  15    & $5/2^-_2 \rightarrow 
9/2^-_3$ & 100 \\
$9/2^-_1 \rightarrow 11/2^-_1$  &  10    & $7/2^-_2 \rightarrow 
11/2^-_4$ & 151 \\
$11/2^-_1 \rightarrow 13/2^-_2$ &  10    & &\\
$1/2^-_1 \rightarrow 5/2^-_1$   & 415    & & \\
$3/2^-_1 \rightarrow 7/2^-_1$   & 353   && \\
$5/2^-_1 \rightarrow 9/2^-_1$   & 328   && \\
$7/2^-_1 \rightarrow 11/2^-_1$  & 308   && \\
$9/2^-_1 \rightarrow 13/2^-_2$  & 301   && \\
\hline
\end{tabular}
  \label{er167-be2} }
\end{table}

Table \ref{er167-be2} shows the B(E2) intra- and inter-band transition
strengths for $^{167}$Er, in units of $e^2b^2 \times 10^{-2}$.
Effective charges were 1.3$e$ and 2.3$e$. Experimental data are shown with
error bars in the figure and in parenthesis in the table.
As usual, the intra-band transitions are in
general two orders of magnitude larger than the inter-band transition
strengths. In both cases the agreement with experiment is very good.

The results shown above faithfully display the usefulness of the pseudo SU(3)
model in the description of normal parity bands in heavy deformed nuclei.
However, as already mentioned, the role of nucleons in intruder parity orbitals
cannot be underestimated. The quasi SU(3) symmetry offers the possibility
to describe them in similar terms as those occupying normal parity orbitals.

\section{Quasi SU(3) symmetry}

The ``quasi SU(3)'' symmetry, uncovered in realistic shell-model
calculations in the {\it pf}-shell, describes the fact that in the case of
well-deformed nuclei the quadrupole-quadrupole and spin-orbit interactions
play a dominant role and pairing can be included as a perturbation.
In terms of a SU(3) basis, it is shown that the ground-state band
is built from the $S=0$ leading irrep which couples
strongly to the leading $S=1$ irreps in the proton and neutron subspaces.
Furthermore, the quadrupole-quadrupole interaction was found to give
dominant weights   to the so-called ``stretched'' coupled representations,
which supports a strong SU(3)-dictated truncation of the model space.

The interplay between the quadrupole-quadrupole and the spin-orbit
interaction has been studied in extensive shell-model calculations\cite{Zuk95}
as well as in the SU(3) basis\cite{Var98}.
In the former case\cite{Zuk95} the authors studied
systems with four protons and four neutrons in the {\em pf} and
{\em sdg} shells, and compared the mainly rotational spectra obtained in a
full space diagonalization of the realistic KLS interaction with those
obtained in a truncated space and a Hamiltonian containing only
quadrupole-quadrupole and spin-orbit interactions. They found that for
realistic values of the parameter strengths the overlap between the states
obtained in the two calculations is always better than (0.95)$^2$. They
also found that while additional terms in the interaction change the
spectrum, the wave functions remain nearly the same, suggesting that the
differences can be accounted for in a perturbative way.

Following these ideas, a truncation scheme suitable for calculations
in a SU(3) basis was worked out\cite{Var98}. In contrast with what was
done in\cite{Jut98}, systems with both protons and neutrons  were
analyzed, and  the interplay of the quadrupole-quadrupole and spin-orbit
interactions was emphasized, while the pairing interaction was not
included in the considerations for building the Hilbert space.

The SU(3) strong-coupled proton-neutron basis span the complete
shell-model space, and  represents an alternative way of enumerating it.
In order to define a definite truncation scheme that is meaningful for deformed
nuclei, in\cite{Var98} we investigated the Hamiltonian
\begin{equation}
H = -{\frac \chi 2}~ Q \cdot Q ~ - ~
C ~\sum_i \vec l_i \cdot \vec
s_i
\label{ham}
\end{equation}
%
where
\begin{equation}
Q = Q^\pi
+ Q^\nu; \hspace{1cm}
Q^{\pi (\nu)}_\mu = \sqrt{\frac {16 \pi} 5} \sum^{Z
(N)}_i r_i^2
Y_{2\mu}(\theta_i,\phi_i)
\end{equation}
%
is the
total mass quadrupole operator, which is just the sum of the proton
($\pi$) and neutron ($\nu$) mass quadrupole terms, restricted
to work within one oscillator shell, and $\vec l_i, \vec s_i$
are the orbital angular momentum and spin of the $i$-th
nucleon, respectively.
An attractive quadrupole-quadrupole Hamiltonian classifies these basis
states according to their $C_2$ values, the larger the $C_2$ the lower
the energy.
The spin-orbit operator can be written as \cite{Var98}
\begin{equation}
~\sum\limits_i ~\vec l_i \cdot \vec s_i =
- {\frac 1 2} \left[ {\frac {(\eta + 3)!} {2 (\eta -1)!}} \right]^{1/2}
    \left[ a^\dagger_{(\eta,0)1/2} \tilde a_{(0,\eta)1/2} \right]
^{ (1,1) L=S=1, J=0}
\end{equation}

Results for $^{22}$Ne are presented in Table \ref{wf-22ne}. Modern shell-model
calculations\cite{Ret90} exhibit more mixing of SU(3) irreps than
previous ones\cite{Aki69}. The ground-state band, often described
as a pure (8,2) state, has important mixing with the spin 1 (9,0) irrep.
The $J = 1$ state with dominant (6,3) $L=1$, $S =0$ components  mixes
strongly with $(7,1)$ $S= 0$ and others, in agreement with the shell-model
results.

\begin{table}[h!]
\tbl{Comparison of the main components of calculated wave functions
for $J = 0$ and 1 states of $^{22}$Ne}
{\footnotesize
\begin{tabular}{|c|c|c|c|c|c|c|}
\hline
  J & ${(\lambda ,\mu )}$
&$L$ & $S$ &$ Q\cdot Q + \vec l\cdot \vec s$
&Ref.\cite{Ret90} &Ref.\cite{Aki69} \\
\hline
0 & (8,2) & 0 & 0 & 54 & 44  & 71 \\
   &
(9,0) & 1 & 1 & 23 & 14 &  5  \\
   & (6,3) & 1 & 1 &  8 &    &  4  \\
   &
(7,1) & 1 & 1 & 11 &  9 &  2  \\
\cline{1-7}
\multicolumn{4}{|c|}{Sum} &
96 & 67 & 82  \\
\cline{1-7}
\multicolumn{7}{|c|}{}
\\ \cline{1-7}
1 &
(6,3) & 1 & 0 & 25 & 31 &  \\
   & (6,3) & 1 & 1 & 11 &  &  \\
   & (6,3) & 2
& 1 &  8 &  &  \\
   & (7,1) & 1 & 0 & 13 & 13 &  \\
   & (7,1) & 1 & 1 &  9
&  &  \\
   & (7,1) & 2 & 1 &  9 &  &
\\\cline{1-7}
\multicolumn{4}{|c|}{Sum} & 75 & 44 &   \\

\hline
\end{tabular} }
\label{wf-22ne}
\end{table}

Extensive calculation of the energy spectra and electromagnetic transitions in
many even-even, even-odd and odd-odd nuclei along the sd-shell\cite{Var02}
confirm that the quasi S(3) symmetry can be used as a useful truncation scheme
even when the spin-orbit splitting is large.

\section{Summary and Conclusions}

A quantitative microscopic description of  normal parity bands and their
B(E2) intra- and inter-band strengths in many even-even and odd-mass
heavy deformed nuclei has been obtained using a realistic Hamiltonian and a
strongly truncated pseudo SU(3) Hilbert space, including in some cases
pseudo-spin 1 states.

In light deformed nuclei the interplay between the 
quadrupole-quadrupole and spin-orbit
interactions can be described in a Hilbert space built up with the
leading S=0 and 1 proton and neutron irreps, in the stretched SU(3) coupling.
In heavy deformed nuclei this quasi SU(3) truncation scheme
will allow the description of nucleons occupying intruder 
single-particle orbits.

Using the {\em pseudo + quasi SU(3)} approach
it should be possible to perform realistic shell-model calculations for
deformed nuclei throughout the periodic table.

\end{document}